\documentclass[]{spie}  %>>> use for US letter paper
\usepackage{amsmath}
\usepackage[]{graphicx}
\usepackage{cite}
\usepackage{float}
\usepackage{enumerate}

\title{ShaneAO: wide science spectrum adaptive optics system for the Lick Observatory} 

\author{Donald Gavel, Renate Kupke, Daren Dillon, Andrew Norton, Chris Ratliff, Jerry Cabak, Andrew Phillips, Connie Rockosi, Rosalie McGurk, Srikar Srinath, Michael Peck, William Deich, Kyle Lanclos, John Gates, Michael Saylor, Jim Ward, Terry Pfister
\skiplinehalf
University of California Observatories, Santa Cruz, CA
}

\authorinfo{Send correspondence to D.G.: gavel@ucolick.org}
\begin{document} 
  \maketitle

\begin{abstract}
A new high-order adaptive optics system is now being commissioned at the Lick Observatory Shane 3-meter telescope in California. This system uses a high return efficiency sodium beacon and a combination of low and high-order deformable mirrors to achieve diffraction-limited imaging over a wide spectrum of infrared science wavelengths covering 0.8 to 2.2 microns. We present the design performance goals and the first on-sky test results. We discuss several innovations that make this system a pathfinder for next generation AO systems. These include a unique woofer-tweeter control that provides full dynamic range correction from tip/tilt to 16 cycles, variable pupil sampling wavefront sensor, new enhanced silver coatings developed at UC Observatories that improve science and LGS throughput, and tight mechanical rigidity that enables a multi-hour diffraction-limited exposure in LGS mode for faint object spectroscopy science.  
\end{abstract}

%>>>> Include a list of keywords after the abstract 


\section{INTRODUCTION}
\label{sec:intro}  % \label{} allows reference to this section

The Lick Observatory has been a pioneer in the application of adaptive optics for astronomical observing, and has had a sodium laser guidestar system operating at the observatory's 3-meter telescope since 1996\cite{Max1997}.  In early 2011 we reported  plans for the construction of a new adaptive optics instrument and laser\cite{Gavel2011}.  The new AO system and its science camera have now been completed and installed at the telescope where it is presently undergoing its on-sky commissioning activities and being used in shared-risk science runs. In this paper we describe the as-built system and report on its early commissioning results.

The overall program goal is to produce a flexible use facility AO instrument that enables high-Strehl imaging and spectroscopy over a range of science wavelengths from the near-infrared and into the visible ($\lambda 0.7 \mu m - 2.2 \mu m$). It will operate in both laser guide star (LGS) and natural guide star (NGS) mode, and function well over a variety of seeing conditions and guide star brightness situations. As presently configured, system has coverage from $\lambda = 1.1 \mu m$ to $2.2 \mu m$, and has the hooks built in to enable a simple upgrade to extend the shorter end of the wavelength range. Initial tests are showing the instrument meets its Strehl performance predictions over the operating wavelength range.

During the first commissioning runs the AO system and science imager were tested on sky in the various operating modes. For AO these include LGS-sensing, NGS-sensing, a choice of wavefront spatial sampling, off-axis tip/tilt sensing in LGS mode, off-axis wavefront sensing in NGS mode. The science camera has been tested for plate scale, field of view, proper operation of filter selection mechanisms, and detector readout modes, in imaging, spectroscopy, and coronagraph configurations.

The AO system is mounted at the cassegrain focus of the telescope, which means it must operate under changing gravity load. This is particularly challenging for adaptive optics opto-mechanical design since all the flexure tolerance requirements scale with the ratio of diffraction limit to seeing size. The objective is to keep the non-common path pointing differences between science camera and tip/tilt sensor within the diffraction limit during long exposures, possibly hours. Key optical mounts are designed particularly stiff and the optical bench is held in a stress-free manner.\cite{Ratliff2014} The structure is designed to keep the nominal mechanical drift low and elastically repeatable so a flexure compensation model can take over to reach the final goal.

In the laboratory integration and test phase it is impossible to test all operating aspects of the system that will be encountered at the telescope. We encountered some interesting problems and for some of these have produced innovative solutions that are worth reporting on. Accurate off-axis field-steers and focus tracking in laser guide star mode are still work in progress. A new and simple approach to wavefront sensing with very dim guide stars has been surprisingly successful. Some signal processing resolved the problem of trying to find a faint tip tilt guide star against the sky background after field steering.

\section{INSTRUMENT DESCRIPTION}

The ShaneAO adaptive optics system with Shane Adaptive Red Camera and Spectrometer (ShARCS) science instrument is a diffraction-limited imager, spectrograph, coronagraph, and polarimeter for the visible and near infrared science bands. Adaptive optics corrects for the nominally $\sim$1 arcsecond seeing blur over an approximately 20 arc second field of view, roughly matching the expected isoplanatic patch\cite{Fried1976} in the longest science wavelength band. With the laser guidestar, the system has full sky coverage with performance dependent on proximity of a natural tip/tilt star, airmass, and seeing conditions.

\begin{figure}
   \begin{center}
   \begin{tabular}{c}
   \includegraphics[height=7cm]{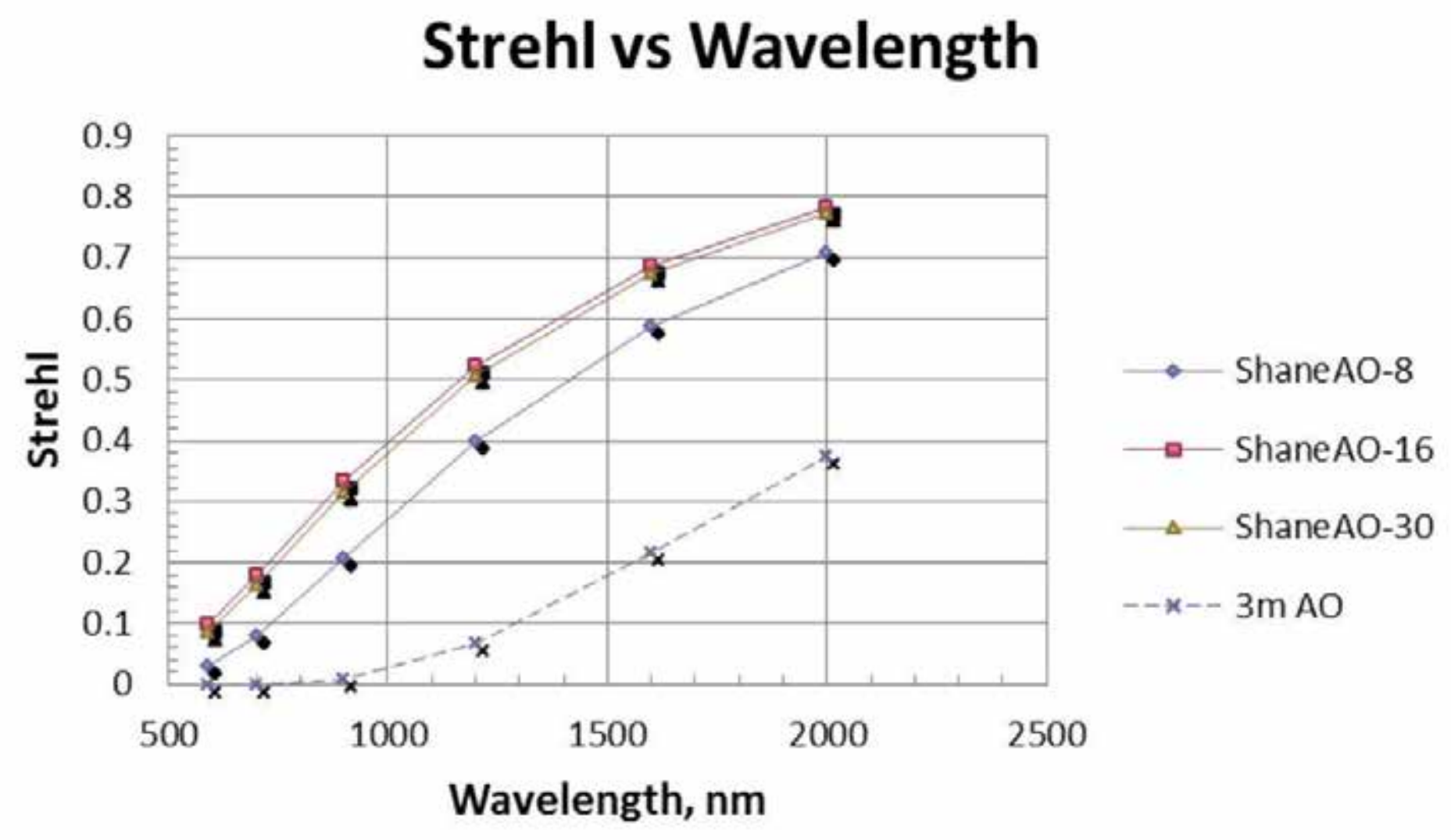}
   \end{tabular}
   \end{center}
   \caption[Predicted] 
%>>>> use \label inside caption to get Fig. number with \ref{}
   { \label{fig:strehl} 
Predicted Strehl performance of ShaneAO and comparison to the older AO system at Lick.}
   \end{figure}

\noindent Unique aspects of the new system from a science prospective include:\cite{McGurk2014, Srinath2014}
\begin{enumerate}[\hspace{12 pt}$\bullet$]\itemsep-4pt
\item {Detector sampling: 0.035 arcseconds per pixel}
\item {Science detector: Hawaii2RG, 80\% QE in the near IR, 70\% 0.7 to 1.0 $\mu$m}
\item {On-axis imaging Strehl: see Figure \ref{fig:strehl}}
\item {Long-exposure imaging stability: hold to diffraction-limit for one hour (goal after flexure model)}
\item {Spectral resolution: R=500 grism (may be upgraded in the near future to R=2000)}
\item {Slit width: 0.1 arcseconds}
\item {Slit angle on sky: adjustable using Cassegrain rotator}
\item {Long-exposure spectroscopic stability: hold to 1/2 slit width for 4 hours (goal after flexure model)}
\item {Minimum brightness natural guide star: $m_v$=13}
\item {Minimum brightness tip/tilt guide star: $m_v$=18}
\item {Infrared sensitivity: $m_K=20$ to $m_J=23$}
\item {Camera readout modes: Correlated double-sampling (CDS), up the ramp (UTR), sub-frame region of interest (ROI), quick take}
\item {Exposure support: Multiple frame co-added, automated nod and expose coordinated with telescope (snap-i-diff, box-4, box-5), automated darks sequence based of history of science exposures}
\item {Observations support: automatic data logging, automatic data archiving}

\end{enumerate}
\noindent Unique aspects from an engineering prospective include:
\begin{enumerate}[\hspace{12 pt}$\bullet$]\itemsep-4pt
\item {Woofer-tweeter deformable mirror pair\cite{Gavel2014}}
\item {Wide-field acquisition of tip/tilt star in LGS mode}
\item {Partially corrected tip/tilt star}
\item {Selectable Hartmann subaperture sampling depending on seeing and wavefront guide star brightness}
\item {Extremely stiff opto-mechanics to enable long-exposure work, particularly narrow-slit (0.1 as) spectroscopy}
\item {Cold-baffled infrared camera, along with cold-stop alignment aids, to minimize extraneous background}
\item {Ultra-reflective broadband coatings for optics in the science and wavefront sensor paths\cite{Phillips2014}}
\end{enumerate}

The AO system with infrared camera represent two of the three components forming the new AO system for the Shane. The third, yet to come, is a new guide star laser. Plans are to later this year incorporate a solid-state laser that was designed and built under the auspices of the Center for Adaptive Optics.\cite{Dawson2006} This laser has a pulse and spectral format that will take maximum advantage of sodium optical pumping\cite{Gavel2012, Rampy2012}. In the mean time, the ancient dye-laser, originally commissioned in 1996\cite{GavelFriedman1998}, will continue to be used as the ShaneAO LGS beacon source throughout the remainder of this year.

%-------------
%-------------

\begin{figure}
   \begin{center}
   \begin{tabular}{c}
   \includegraphics[height=7cm]{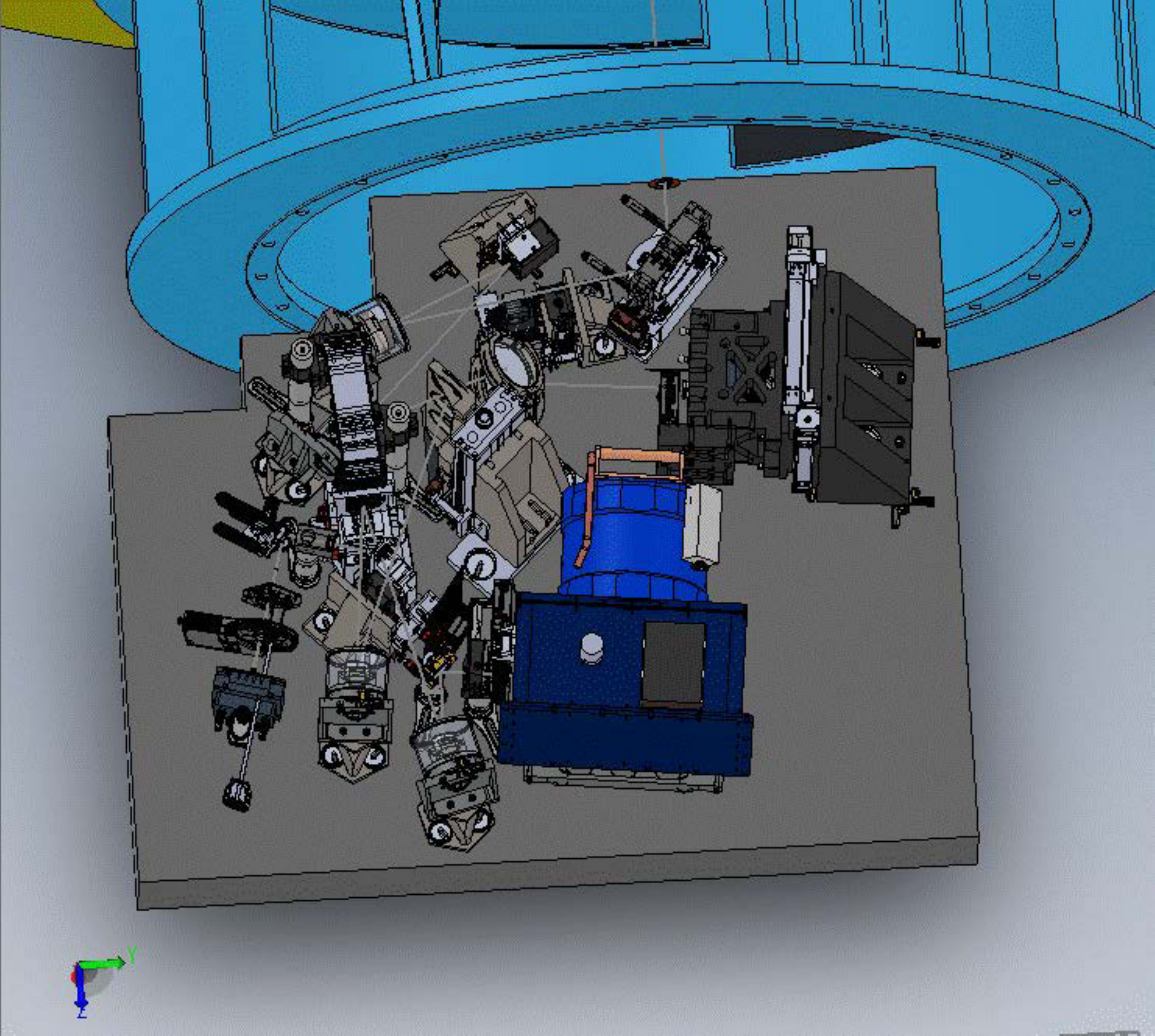}
   \includegraphics[height=7cm]{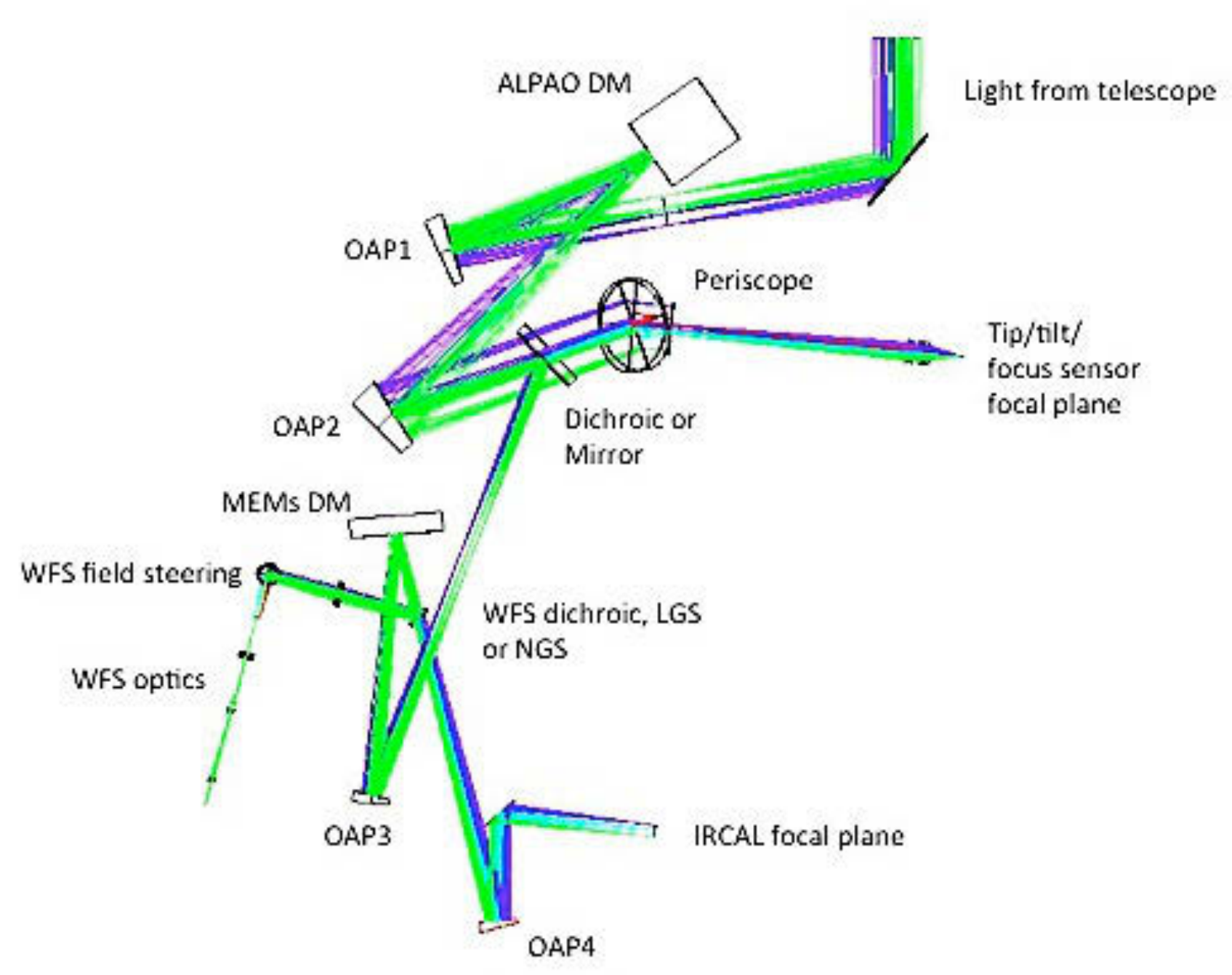}
   \end{tabular}
   \end{center}
   \caption[example] 
%>>>> use \label inside caption to get Fig. number with \ref{}
   { \label{fig:layout} 
Computer renditions of the optomechanical layout(left) and optical paths (right) for ShaneAO.}
   \end{figure}

The optical table layout is shown in Figure \ref{fig:layout}. The AO optics and the ShARCS dewar are all mounted on a single optical bench, which is oriented vertically when the telescope is pointed to zenith. Light from the telescope enters  at the top of the bench with a focal ratio of approximately f/17.5. This light is collimated with a pupil conjugate relayed to the first deformable mirror, which is the woofer. The woofer mirror corrects low order aberrations, including tip/tilt, at high stroke. It is an ALPAO 52-element magnetically actuated membrane mirror with up to $\pm 50 \mu m$ stroke. The use of the membrane mirror to handle the fast beam steering is a unique feature of the ShaneAO system. This avoids having an additional tip/tilt mirror or having to mount the DM on a fast steering stage. The woofer DM surface deflection in tilt corresponds to a $\pm$4 arcseconds range on sky. It was a concern at first that wind shake or poor telescope tracking might saturate this range. However, the Shane telescope has little to no resonant vibration (data reported in our 2011 paper\cite{Gavel2011}) and the improved telescope tracking system, aided with periodic tilt offloads from the AO system, has proven quite capable of keeping tip/tilt out of saturation.

%-------------
%-------------

After the woofer, the light is split dichroically, sending $600$ nm $< \lambda < 900$ nm light on a 120 arcsecond field to the tip/tilt sensor and the central 20 arcsecond field on to the second relay. The tip/tilt sensor therefore sees partially corrected starlight, corrected by the woofer DM. This PSF correction of the tip/tilt star is readily apparent in the real-time display of the tip/tilt sensor, noticeably sharpening and brightening it. This sharpening improves the tip/tilt centroiding sensitivity, which increases the limiting magnitude with a consequence of improved sky coverage in LGS mode. It is actually better to only low-order correct this star, as the high-order corrections are degraded by isoplanatism and since LGS tip/tilt stars are in most cases well outside the isoplanatic patch, attempting to add the high-order wavefront correction will only degrade the tip/tilt star PSF.

The second relay is a narrow-field (20 arcsecond) one, with the pupil conjugated at a MEMS (MicroElectro-Mechanical System) deformable mirror. The MEMS is a 1020-actuator KILO-DM produced by Boston Micromachines Corporation. Per our specification, this device has zero defective actuators inside the illuminated beam footprint on the mirror. At roughly 30-actuators across the beam, this actually super-samples the pupil relative to the two choices of Hartmann sampling in ShaneAO's wavefront sensor (described below). With a non-MEMS device this would have presented a problem; the extra degrees of freedom could drift into unobservable aberration modes. The MEMS device however has excellent go-to repeatability with negligible hysteresis, and with a careful construction of the real-time control matrix, the unobservable modes are avoided. Deep measurements of the closed-loop PSF show no artifacts induced by unobservable modes. The PSF of a MEMS however does show a characteristic MEMS diffraction pattern, with regular patterns of horizontal and vertical dots at $\lambda$ / spacing of the actuators, due to an actuator-post ``print through'' that is common to these devices.

The input of the first relay is matched to the focal ratio of the telescope at approximately f/17.5. The output of the first relay is f/28.5.  This is transferred 1:1 by the second relay, into the science camera. The beam going into the tip/tilt sensor is converted to f/5 at the tip/tilt sensor, making its plate scale 0.3 arcsecond per pixel and providing it with a field of view of 19 arcseconds. This does not cover the entire tip/tilt star field, so the sensor head (along with f/5 lens) is mounted on an x-y translation stage with enough travel to cover the field. The tip/tilt sensor is a Marconi CCD39 chip, 80 $\times$ 80 pixel, with 6 electrons read noise at up to 1kHz frame rate. SciMeasure Analytical Systems produced the camera controller.

The Shack-Hartmann wavefront sensor accepts a collimated beam split after the MEMS deformable mirror (hence the system runs in closed loop). The splitter sends light at $\lambda < 1\mu$m. This includes the laser guidestar return at $\lambda = 589$nm in LGS mode, or starlight at $\lambda < 900$nm in NGS mode. In NGS mode, the dichroic in the first relay is replaced by a mirror to capture for the wavefront sensor what would have gone to the tip/tilt sensor. The wavefront sensing beam is focused through a field stop and recollimated on to a lenslet array whose focused dots are then relayed to the wavefront sensor camera. Two sets of Hartmann collimator/lenslet/relay assemblies can be switched in to perform the wavefront measurement. The low-resolution mode is 8 lenslets across the pupil (40 cm subapertures at the primary), which is useful for sensing dim natural guide stars and the present dye laser return in modest sodium density conditions. The high-resolution mode is 15 lenslets across (20 cm subapertures at the primary) (for historical reasons called 16x mode), which is useful for producing higher Strehls at the shorter science wavelengths but would need a brighter guidestar. We expect the new fiber laser to have enough return signal to routinely run the AO system in 16x mode.

The field stop at the wavefront sensor focus functions to block the Rayleigh backscatter (scattered light from the air molecules at low altitude) from fogging the wavefront sensor. This is a fixed opening that is sized for the field of the subapertures on the wavefront sensor CCD array, approximately 6 arcseconds diameter in this design. The opening is in the center of a wide blackened disk which acts to further suppress the Rayleigh light from scattering back in to the sensor. The stop has proven during the commissioning runs to be extremely effective in blocking the Rayleigh; in most cases the Rayleigh background was not even detectable above sky background. This is a huge improvement over the old system with its flexy motorized iris and and positioning stages. It also has an important cost benefit: it avoids requiring a Rayleigh-blanking laser pulse format and saves all the system complexities associated with gating the return.

The wavefront sensor detector is a Lincoln Laboratories CCID66 array, 120x120 pixels with 1-2 electrons read noise at up to 1.5 kHz frame rate. This camera controller is also built by SciMeasure Analytical Systems.

The science camera is based around a Hawaii2RG 2k$\times$2k pixel infrared detector array. The camera system description and its performance report is given in a companion paper \cite{McGurk2014}. The detector is an engineering grade A device, in our case having about 600 inoperative rows, but otherwise over the rest of the array meeting science grade requirements. The usable portion of the array provides plenty of detector real-estate to fit the 20 arcsecond diameter AO science field, and allows enough room for dispersed spectra as well, even for the envisioned future upgrade to higher spectral resolution capability.

The science beam is collimated within the cryogenic dewar where it is transmitted through filters and a cold pupil stop. The cold stop, located at a pupil conjugate, blocks stray infrared light coming from paths other than directly from the sky. The positioning of this cold stop and the alignment of the beam into it was a problem in the prior IRCAL instrument. In the new camera one of the filter wheel positions contains a pupil viewing lens (imaging the pupil on to the detector) which greatly helped the alignment process.  Careful adjustments during the instrument assembly placed the cold stop in the correct location. The pupil viewing lens then allowed aligning the AO feed beam precisely on axis to the cold stop. The result is that sensitivity is greatly improved due to reduced background noise in the focal plane. This appears to be meeting our prediction of a 12x improvement over the old system for point source exposure time to a given signal-to-noise (Section 3 in the companion paper on ShARCS\cite{McGurk2014}).

\section{EARLY COMMISSIONING RESULTS}

First light commissioning runs took place 11-15 April 2014 and 6-11 May 2014. Of these 11 nights, 6 had the laser available. These were followed by 12 nights of shared-risk science observing of which 7 used the laser. The laser takes considerable additional effort since it requires a laser operator and two plane spotters, plus extra activities of preparing and communicating laser pointing lists for satellite avoidance clearance and notifying the FAA. The observatory is short on budget and manpower so although most AO observers would prefer to use the laser, only about half of those requests can be accommodated this year. The demand going forward for AO instrument on the Shane is anticipated to be at roughly this 33\% of available nights level.

\subsection{Bright guide stars}

\begin{figure}[H]
   \begin{center}
   \begin{tabular}{c}
   \includegraphics[width=16cm]{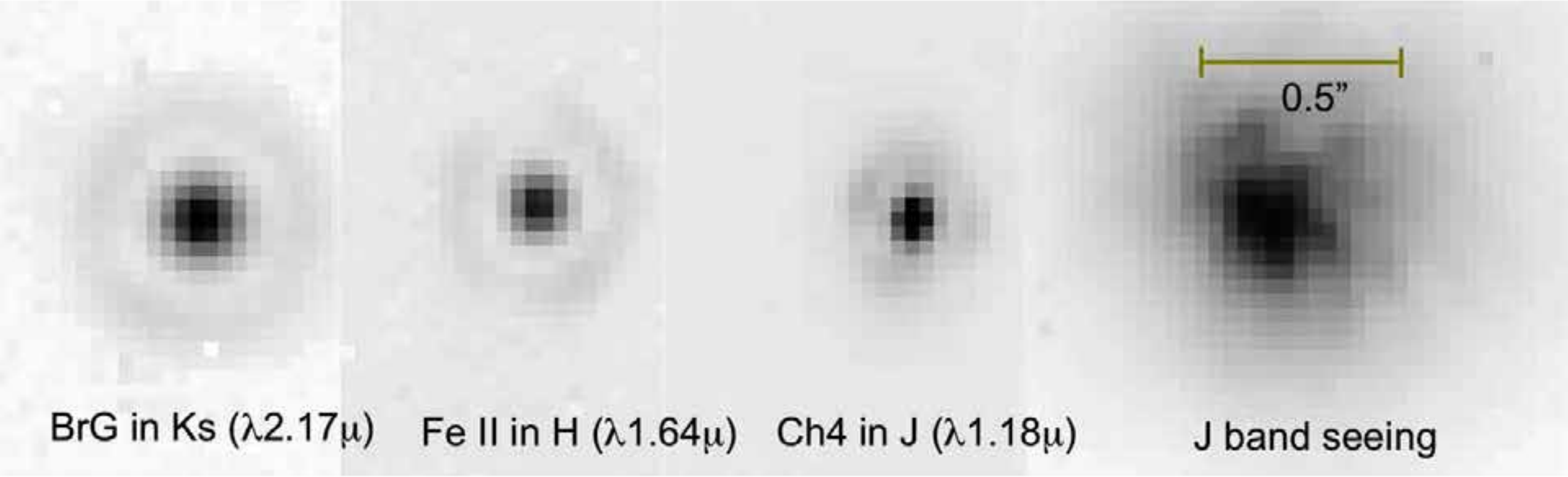}
   \end{tabular}
   \end{center}
   \caption[example] 
%>>>> use \label inside caption to get Fig. number with \ref{}
   { \label{fig:psf} 
Point spread functions at each of the science bands.}
   \end{figure}

\begin{figure}[H]
   \begin{center}
   \begin{tabular}{c}
   \includegraphics[width=10cm]{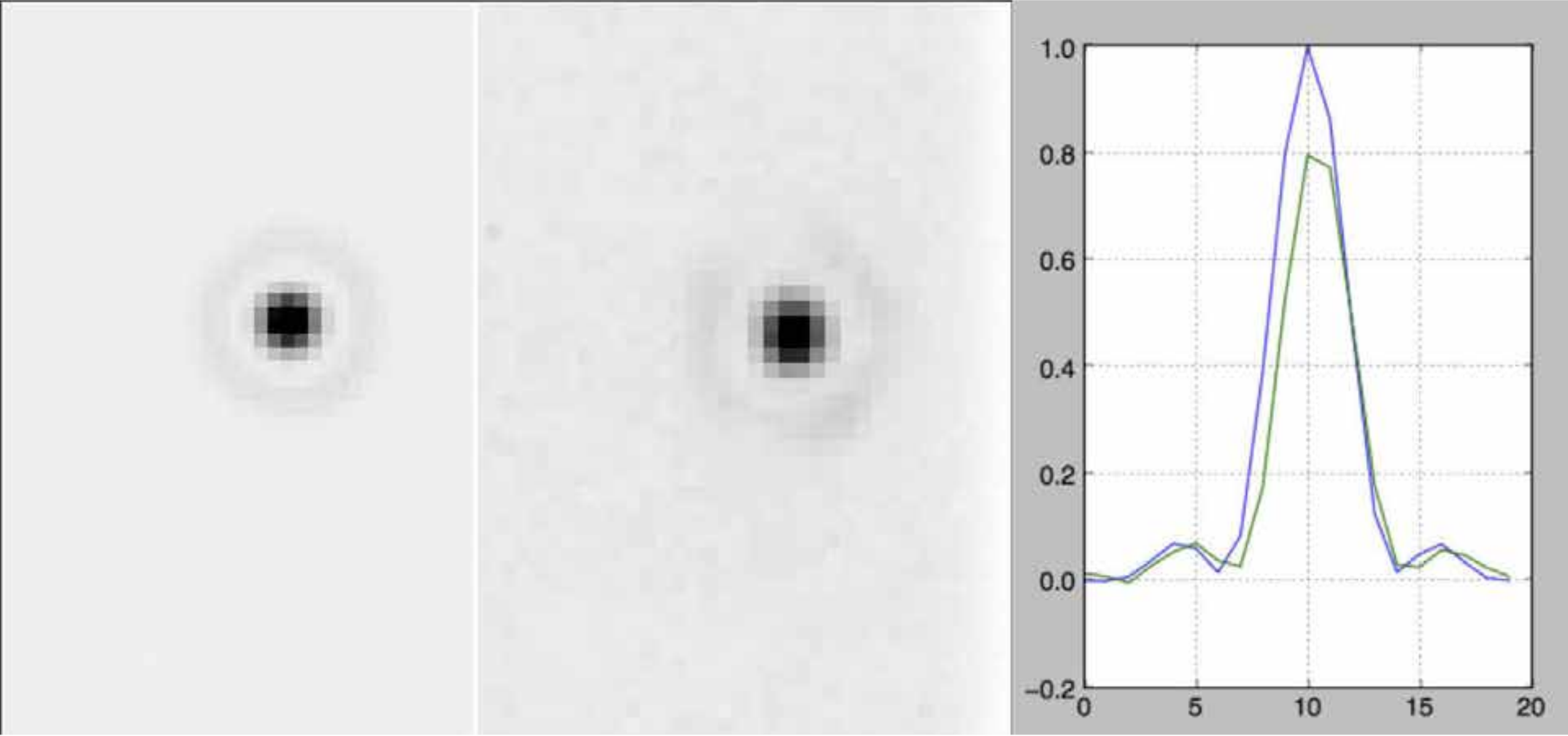}
   \end{tabular}
   \end{center}
   \caption[example] 
%>>>> use \label inside caption to get Fig. number with \ref{}
   { \label{fig:psfb} 
Left: PSF of internal source in H band (image-sharpened). Center: AO corrected H band PSF on sky. The on-sky Strehl is approximately 0.8 relative to the calibrator (lineout comparison on right).}
   \end{figure}

\begin{figure}[H]
   \begin{center}
   \begin{tabular}{c}
   \includegraphics[height=5cm]{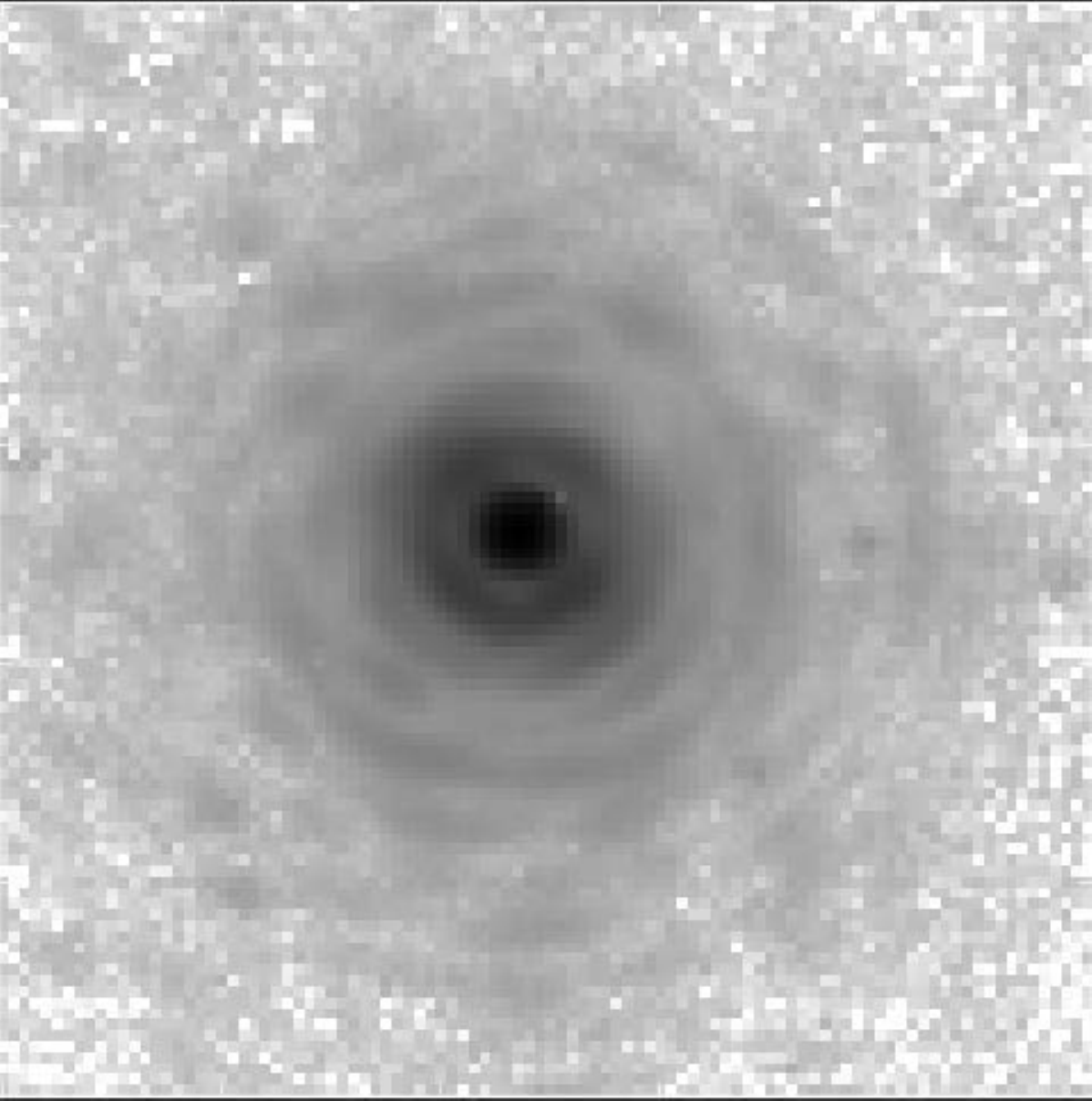}
   \end{tabular}
   \end{center}
   \caption[example] 
%>>>> use \label inside caption to get Fig. number with \ref{}
   { \label{fig:AiryRings} 
Log-stretch of the K band PSF, showing the Airy rings. The brightening of approximately every fourth ring is due to diffraction of the 80 cm secondary obscuration mixing with diffraction of the Shane 3.05 meter primary.}
   \end{figure}

AO on bright natural guide stars worked quite well immediately. First light PSFs show multiple Airy rings and high Strehl at all the targeted science wavelength bands (Figures \ref{fig:psf}, \ref{fig:psfb} and \ref{fig:AiryRings}). The Strehl ratios lie on or above our predicted Strehl curves, with the caveat that the prediction curves include the entire error budget using a laser guidestar with 100 photons/s/cm2/W return brightness (comparable to what we expect with the new fiber laser) while these initial data are taken on bright natural stars. The bright NGS results demonstrate that there are no unanticipated AO systematic errors preventing achieving the laser results. As mentioned earlier, laser performance measurements await arrival of the new laser later in the year. Cone effect and laser spot elongation are negligible on the 3-meter aperture so we expect the laser return brightness to be the key parameter.

%-------------
%-------------

%-------------
%-------------

%-------------
%-------------

\subsection{Faint guide stars}

On the nights following first light we began investigating performance with fainter guidestars. We found that at low signal levels the AO control started behaving unstably, a consequence of failing to properly account for low intensity in the denominator of the Hartmann centroider equation. The simplified version of this equation is
\begin{equation}
c = {{A - B} \over { A + B + r}}
\end{equation}
where $A$ and $B$ are wavefront sensor pixel counts and $r$ is a regularization term intended to prevent dividing by small numbers when the total intensity is low. For high intensity, the difference signal is normalized by total intensity so that the result depends only on spot position. At low intensity however, the background noise, which is dominated by the sky background, makes this result very erratic unless $r$ is sufficiently high. A detailed statistical analysis, with shot noise from the star, the sky, and accounting for the fact that two random variables (intensity and star position) get multiplied in the process, we would end up with an optimal Weiner solution:
\begin{equation}
r = \left( A + B \right)\times{\rm SNR}^{-2}
\end{equation}
The problem with this answer is that it depends on a random and initially unknown quantity, $A+B$. Instead, we implemented a dumb solution
\begin{equation}
r = \sigma \left({\rm sky background}\right)
\end{equation}
This worked surprisingly well. The AO system was able to lock stably on any brightness star with a degradation in wavefront correction that is commensurate with what is expected with the error budget model. The feedback control remained stable even with no star in the sensor (stable, but obviously with no correction). Although this regularization does not follow the same gain profile with SNR as does the Weiner solution, it has the same desirable qualities: basically transparent (gain of one) at high signal, a degradation to gain of one-half at intensity equal to the sky background standard deviation, and on-average zero gain when there is no signal.

\begin{figure}
   \begin{center}
   \begin{tabular}{c}
   \includegraphics[height=9cm]{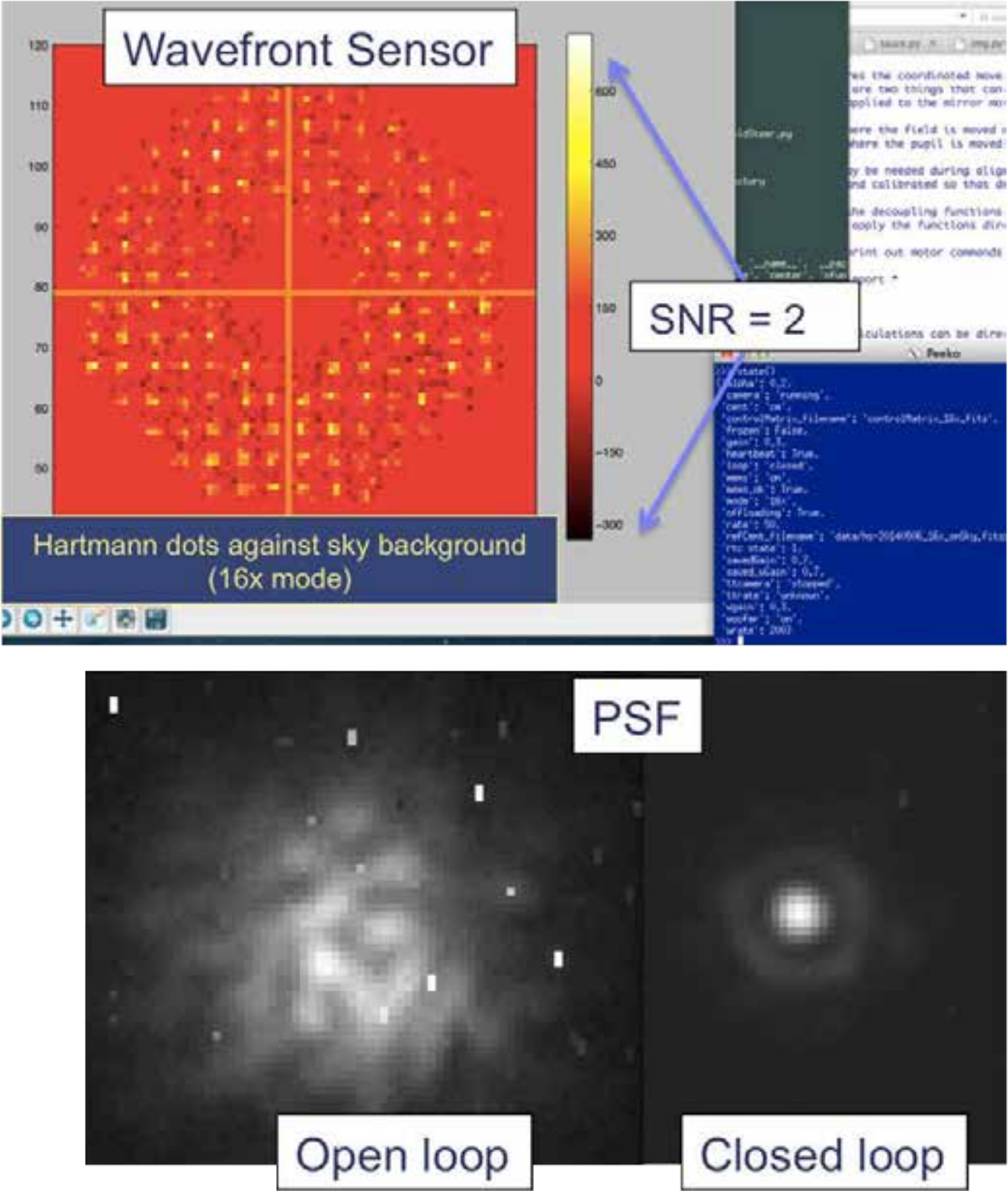}
   \end{tabular}
   \end{center}
   \caption[example] 
%>>>> use \label inside caption to get Fig. number with \ref{}
   { \label{fig:lowSNR} 
Wavefront sensor display and open and closed loop PSFs in K band with an 11'th magnitude natural guide star.}
   \end{figure}

The NGS mode Strehl performance versus guide star magnitude is still undergoing data collection and analysis. Some preliminary results are promising. Figure \ref{fig:lowSNR}  shows that at approximately $m_v = 11$ the signal to noise on the 16x Hartmann sensor running at 50 Hz frame rate is about 2. The corresponding open and closed loop K band PSFs are shown below it.

%-------------
%-------------

\begin{figure}
   \begin{center}
   \begin{tabular}{c}
   \includegraphics[height=5cm]{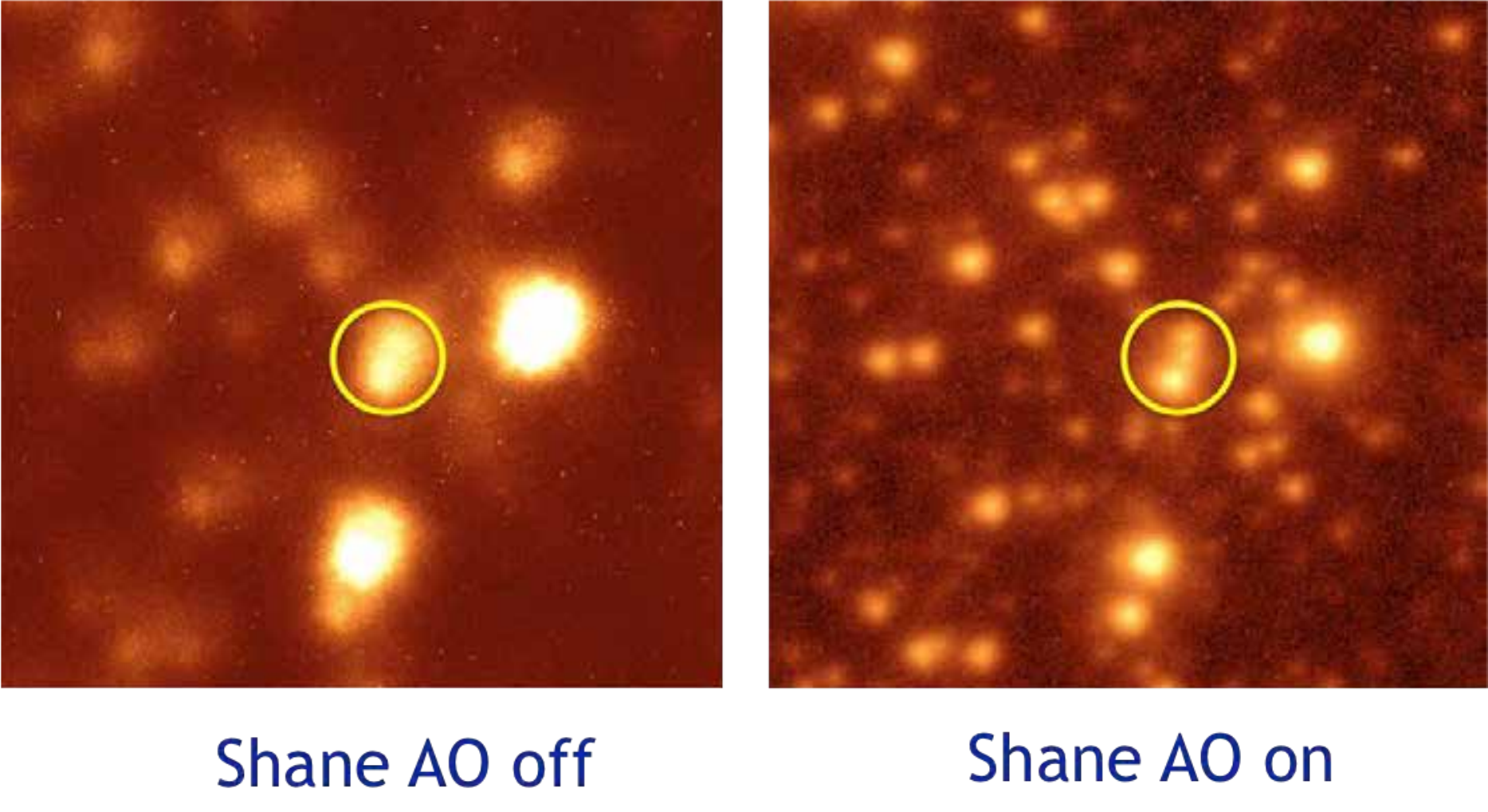}
   \end{tabular}
   \end{center}
   \caption[example] 
%>>>> use \label inside caption to get Fig. number with \ref{}
   { \label{fig:cluster} 
Portion of the M92 globular cluster taken in H band. The stars in the circle, which have a spacing of about 0.4 arcseconds, are blurred together in normal seeing but separated in the AO corrected image.}
   \end{figure}

Figure \ref{fig:cluster} shows the globular cluster M92 in H band. This image that was taken with the AO locked on an 11'th magnitude star in the cluster.

%-------------
%-------------

LGS mode faint tip/tilt guide star performance is discussed in subsection \ref{subsec:fieldSteer} below.

\subsection{LGS mode}

The current dye laser was not expected to produce a guidestar bright enough for use with the 16x wavefront sensor. However, we did successfully close the LGS AO loop in 16x mode at 50 Hz frame rate during the April run, which had two nights of reasonably good seeing. In average or poor seeing, as experienced during the May run, only the 8x wavefront sensor mode could provide good enough signal for AO. Sodium column densities in the early Spring are also historically low, further reducing our ability to complete tests of LGS performance. Using the 8x wavefront sensor keeps ShaneAO's LGS mode on par with the previous AO system, which used a 7 subapertures-across wavefront sensor. This will suffice until the new fiber laser comes on line later this year. The fiber laser should make a significantly brighter guidestar and with this we expect to run in 16x mode routinely.

\begin{figure}[H]
   \begin{center}
   \begin{tabular}{c}
   \includegraphics[width=16.cm]{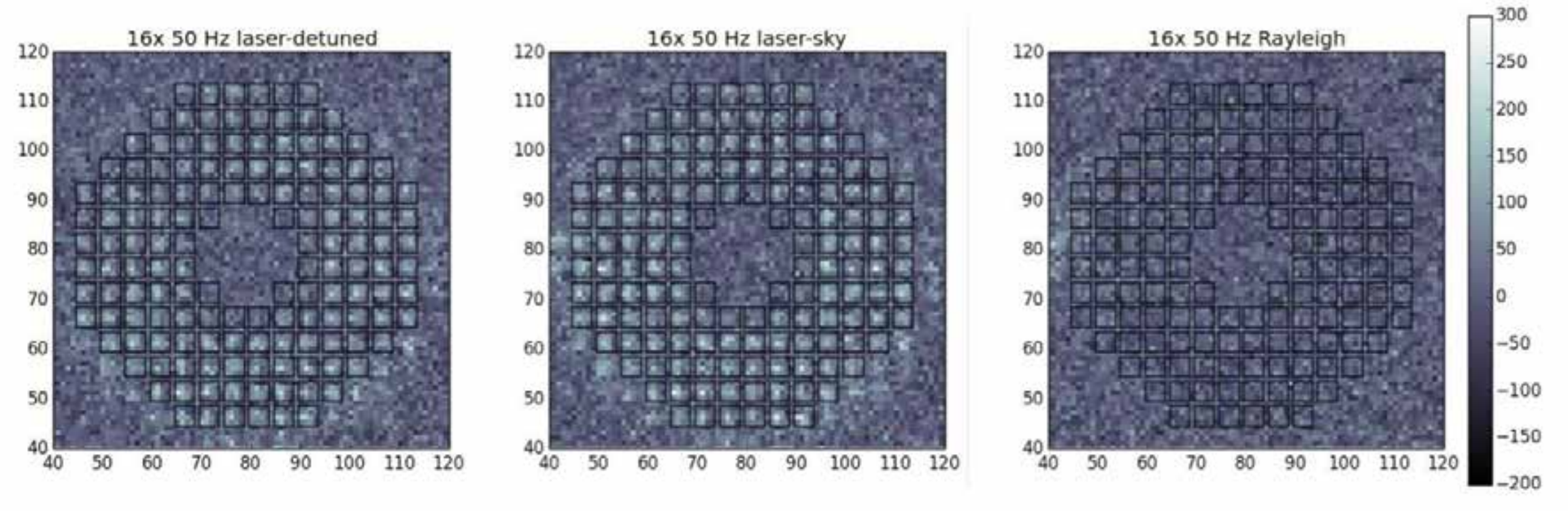}
   \end{tabular}
   \end{center}
   \caption[example] 
%>>>> use \label inside caption to get Fig. number with \ref{}
   { \label{fig:Rayleigh16} 
Laser guide star as seen by the wavefront sensor in 16x mode. Left: LGS minus Rayleigh, Center: LGS minus sky, Right: Rayleigh minus sky.}
   \end{figure}

\begin{figure}[H]
   \begin{center}
   \begin{tabular}{c}
   \includegraphics[width=16.cm]{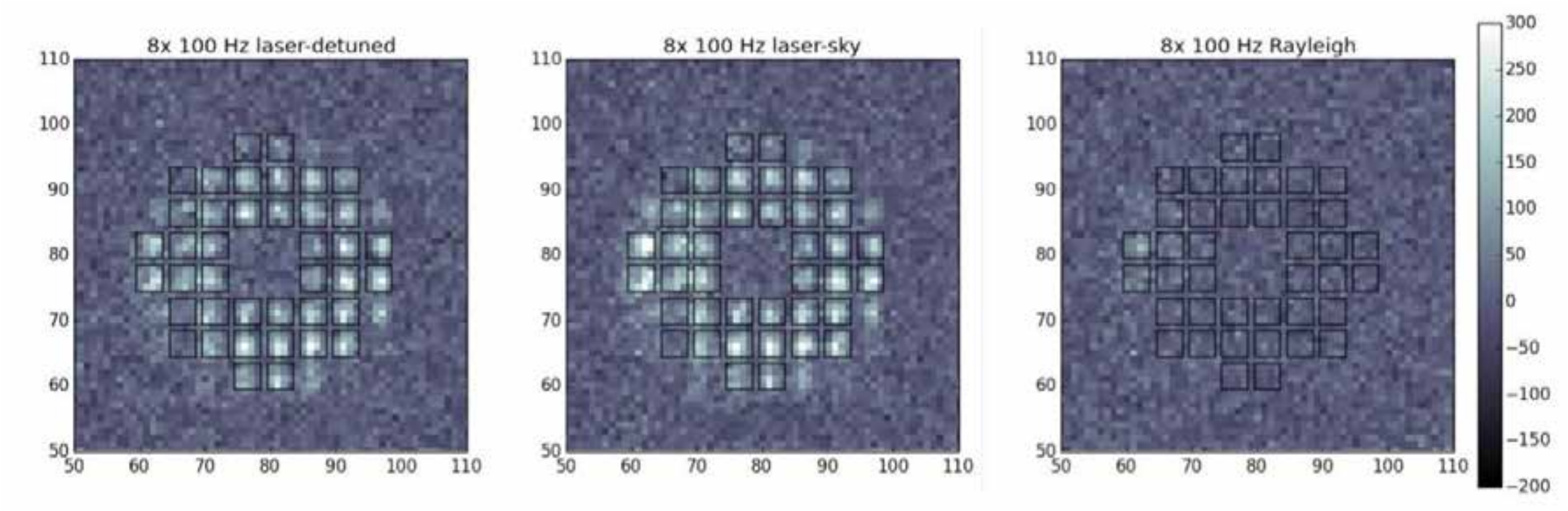}
   \end{tabular}
   \end{center}
   \caption[example] 
%>>>> use \label inside caption to get Fig. number with \ref{}
   { \label{fig:Rayleigh8} 
Laser guide star as seen by the wavefront sensor in 8x mode. Left: LGS minus Rayleigh, Center: LGS minus sky, Right: Rayleigh minus sky.}
   \end{figure}

We performed experiments to measure the amount of added wavefront sensor background due to the backscatter of Rayleigh light. One normally measures the mean background by tuning the laser off the sodium wavelength, rather than shuttering it, in order to account for Rayleigh background. We compared the background averages taken with tuning off versus shuttering and found there to be little difference (Figures \ref{fig:Rayleigh16} and \ref{fig:Rayleigh8}). A field stop at the focal plane in front of the wavefront sensor is designed to block the Rayleigh from entering. During fabrication, considerable care was taken to properly size and position the stop as well as to absorb the unwanted ghost reflections of the Rayleigh light, which can be hundreds of times brighter than the LGS beacon. Our experiments confirmed that the field stop is indeed blocking the Rayleigh light nearly to the level of the sky background. None of the subapertures have to be de-weighted or removed from the wavefront reconstruction as they were in the old system.

%-------------
%-------------

%-------------
%-------------

\subsection{Tip/Tilt sensor field steering, LGS mode}
\label{subsec:fieldSteer} 

In LGS mode, the tip/tilt star is selected from a 120 arcsecond diameter field using a detector mounted on a translation stage. Knowing the relative position of the tip/tilt guidestar target relative to the science object is crucial to accurately positioning the science object on the desired part of the science detector. We performed tests of blind steering to faint guidestars up to 60 arcseconds off axis. Blind steering involves moving the detector stage and then setting the center target on the detector.  It is impossible for an operator to see a faint tip/tilt star against the sky background when displaying individual frames even at the lowest frame rate ($f_s=40$ Hz; $T_s=25$ ms), which impedes setting the reference point by eye.  If the reference point is properly set, the tip/tilt loop can take over. The bandwidth of the tip/tilt control loop, $f_c \sim f_s/10$, acts to filter the noise and tracks the star even when individual frames have signal-to-noise less than one.

To remedy the acquisition problem, we implemented a moving average filter for the display, which aids in bringing the star image above the noise for the operator's eyes. This works quite well to beyond 17'th magnitude. The tip/tilt loop was closed with reasonably low tip/tilt residual on stars as faint as $m_v$16.5. Further testing on sky with revisions to the field steering procedure will determine if we can reach the $m_v$ 18 goal.

\subsection{Wavefront Sensor field steering, NGS mode}

In NGS mode, the natural guide star must be steered into the wavefront sensor. A pair of motorized pointing-and-centering (P\&C) mirrors mounted in a periscope configuration accomplish this task. The P\&C is designed to operate over a 20 arcsecond diameter field for selecting the guide star. The P\&C needs to position the beam accurately on both the focal plane (detector) and the pupil plane (Hartmann lenslet array). Even a 1/10 subaperture shift of the pupil plane results in marginal stability as it offsets the image of the deformable mirror actuator locations relative to assumed locations on the measured wavefront. At extreme field points the pupil image will rotate, i.e. twist about the optical centerline, enough to cause stability issues at high feedback gain.

The field steering software uses a ray trace model to predict the pointing and centering of the beam as a function of P\&C mirror tilts. It then inverts this model to obtain mirror commands corresponding to a desired field steering offset. Unfortunately, the model does not accurately enough predict the physical action of the mirrors for stable centering, so the model will have to be adjusted using empirical data. During the April/May commissioning runs we were able to repeatably steer to a number of calibrated field points and to steer repeatably for offset into the science spectrograph slit. However, a P\&C model that will enable arbitrary NGS offset is still work in progress.

\subsection{Image sharpening}

\begin{figure}
   \begin{center}
   \begin{tabular}{c}
   \includegraphics[width=16cm]{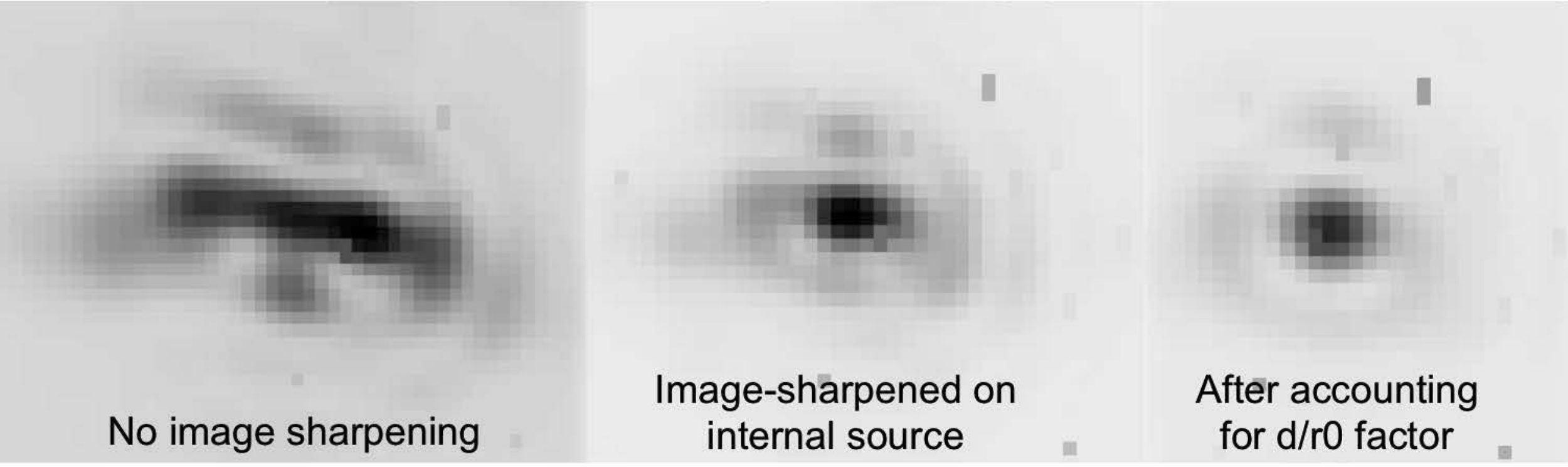}
   \end{tabular}
   \end{center}
   \caption[example] 
%>>>> use \label inside caption to get Fig. number with \ref{}
   { \label{fig:imageSharpen} 
Progress of image sharpening, starting (left) with the original closed loop on-sky PSF in H band. After image-sharpening on the internal source the on-sky PSF looks like the center image. The rightmost image shows the on-sky PSF after applying a $d/r_0$ offset factor that accounts for the seeing-limited size of the natural guide star.}
   \end{figure}

Image sharpening is a means of correcting for the non-common path aberrations between the wavefront sensor and the science camera. In particular, the ShaneAO wavefront sensor does not probe the powered optics inside the ShARCS dewar, and these introduce astigmatism. The process consists of adjusting offsets of the reference values of Hartmann dot centroids. These are driven slightly away from the nominal crosshair-centers of the 4$\times$4 pixel arrays dedicated to each subaperture. The closed-loop PSF on sky when the AO system is locked on these nominal centers appears quite aberrated (left image in Figure \ref{fig:imageSharpen}). Reference centroid offsets are determined while the AO system is locked on the internal calibrator source located at the input focus. Adjustments are made sequentially for each of the first 12 Zernike modes not counting piston tip and tilt (through radial 4th order). Peak intensity in the PSF is the metric for completing each adjustment.

After the image-sharpening adjustments, the PSF of the internal calibrator source looks close to that shown in Figure \ref{fig:psfb}. On-sky however, the PSF is still aberrated. The problem is that the internal calibrator source looks different to the wavefront sensor than the on-sky natural guide star, in particular the NGS is larger in each subaperture by a factor of $d/r_0$ because of seeing. The centroid offsets depend on the Hartmann spot size, so they must be appropriately modified to account for the change in spot size in transferring from internal source to NGS source.
The rightmost image in Figure \ref{fig:imageSharpen} shows the on-sky PSF after accounting for a $d/r_0$ factor.

%-------------
%-------------

\section{NEXT STEPS}

Additional engineering time is scheduled throughout the remainder of this year, along with an increasing allocation of science time to observers. The laser schedule is somewhat uncertain but we expect to take it to the telescope before the end of this year (2014) and to begin on sky experiments with it. It is designed so that we can swap the beam feed at the focus of the laser launch telescope with the old laser, thereby enabling on-sky commissioning of the new laser to proceed initially without disabling use of the old laser for continued science observing.

\noindent A number of clean-up activities continue with the AO system and ShARCS camera. They are:
\begin{enumerate}[\hspace{12 pt}$\bullet$]\itemsep-4pt
\item{Complete the building and testing of a gravity-induced non-common-path flexure model}
\item{Build the nonlinear empirical models for NGS mode field steering}
\item{Add internal baffling to the ShARCS camera and re-measure background and ghosting characteristics}
\item{Enhance operations efficiency with additions to the graphical user interfaces and automated scripts, per suggestions from the instrument operators}
\end{enumerate}

\section{CONCLUSION}

We completed construction of the new Shane adaptive optics system and infrared science camera and performed a major portion of the on-sky commissioning tests during the Spring 2014 semester. The system is performing well, essentially as originally proposed, and represents a major step forward in AO capability for the Observatory. The AO upgrade process will continue with the addition of a sodium return optimizing fiber guide star laser later this year.

%%%%%%%%%%%%%%%%%%%%%%%%%%%%%%%%%%%%%%%%%%%%%%%%%%%%%%%%%%%%%
\acknowledgments     %>>>> equivalent to \section*{ACKNOWLEDGMENTS}       
 This project was funded by the National Science Foundation, Major Research Instrumentation grant \#0923585. The authors gratefully acknowledge the support of the NSF and also support from the University of California Observatories who provided cost-sharing and in-kind contributions to the project.  

%%%%%%%%%%%%%%%%%%%%%%%%%%%%%%%%%%%%%%%%%%%%%%%%%%%%%%%%%%%%%
%%%%% References %%%%%

%\bibliography{report}   %>>>> bibliography data in report.bib
%\bibliographystyle{spiebib}   %>>>> makes bibtex use spiebib.bst

\end{document}